\documentclass [epsf,useAMS,usenatbib,usegraphicx]{mn2e}
\usepackage{epsf}

\def\grb{GRB\,090323}
\def\zb{$z=3.5673$}
\def\zr{$z=3.5774$}

\newcommand{\cm}{\mbox{cm$^{-2}$}}
\newcommand{\kms}{\mbox{km s$^{-1}$}}

\def\spose#1{\hbox to 0pt{#1\hss}}
\newcommand\lsim{\mathrel{\spose{\lower 3pt\hbox{$\mathchar"218$}}
     \raise 2.0pt\hbox{$\mathchar"13C$}}}
\newcommand\gsim{\mathrel{\spose{\lower 3pt\hbox{$\mathchar"218$}}
     \raise 2.0pt\hbox{$\mathchar"13E$}}}

\title[Super-solar Metal Abundances in Two $z\sim 3.57$ Galaxies]{Super-solar Metal Abundances in Two Galaxies at $z\sim 3.57$ revealed by the \grb\ Afterglow Spectrum\thanks{Partly based on observations collected at the European Southern Observatory under ESO proposal No. 082.A-0693.}}
\author[S.\ Savaglio et al.]{S.\ Savaglio$^{1}$\thanks{E-mail:
savaglio@mpe.mpg.de}, A.\ Rau$^{1}$, J.\ Greiner$^{1}$, T.\ Kr\"uhler$^{1,2,3}$, S.\ McBreen$^{4,1}$, D.\ H.\ Hartmann$^{5}$, 
\newauthor A.\ C.\ Updike$^{5}$, R.\ Filgas$^{1}$, S.\ Klose$^{6}$, P.\ Afonso$^{1}$, C.\ Clemens$^{1}$, A. K\"upc\"u Yolda\c{s}$^{7}$, 
\newauthor F.\ Olivares E.$^{1}$, V.\ Sudilovsky$^{1,2}$ and G.\ Szokoly$^{8,1}$\\
$^{1}$Max Planck Institute for Extraterrestrial Physics, 85748 Garching bei M{\"u}nchen, Germany \\
$^{2}$Physics Department, Technische Universit\"{a}t M\"{u}nchen, James Franck Str., 85748 Garching, Germany \\
$^{3}$Dark Cosmology Centre, Niels Bohr Institute, University of Copenhagen, Juliane Maries Vej 30, 2100 Copenhagen, Denmark\\
$^{4}$School of Physics, Science Center North, University College Dublin, Belfield, Dublin 4, Ireland \\
$^{5}$Clemson University,  Department of Physics and Astronomy Clemson, SC 29634, USA \\
$^{6}$Th\"uringer Landessternwarte Tautenburg, Sternwarte 5,  07778 Tautenburg,  Germany \\
$^{7}$European Southern Observatory, 85748 Garching bei M{\"u}nchen, Germany \\
$^{8}$E\"otv\"os University, 1117 Budapest, Pazmany P.stny. 1/A, Hungary}

\begin{document}

\date{Submitted 2011 June. Received ...}

\pagerange{\pageref{firstpage}--\pageref{lastpage}} \pubyear{2002}

\maketitle

\label{firstpage}

\begin{abstract}
We report on the surprisingly high metallicity measured in two absorption systems at high redshift, detected in the Very Large Telescope spectrum of the afterglow of the gamma-ray burst \grb. The two systems, at redshift \zb\ and \zr\ (separation $\Delta v \approx 660$ \kms), are dominated by the neutral gas in the interstellar medium of the parent galaxies.  From the singly ionized zinc and sulfur, we estimate oversolar metallicities of [Zn/H] $=+0.29\pm0.10$ and  [S/H] = $+0.67 \pm 0.34$, in the blue and red absorber, respectively. These are the highest metallicities ever measured in galaxies at $z>3$. We propose that the two systems trace two galaxies in the process of merging, whose star formation and metallicity are heightened by the interaction. This enhanced star formation might also have triggered the birth of the GRB progenitor. As typically seen in star-forming galaxies, the fine-structure absorption Si\,\textsc{ii}$^*$ is detected, both in G0 and G1. From the rest-frame UV emission in the GRB location, we derive a relatively high, not corrected for dust extinction, star-formation rate SFR $\approx 6$ M$_\odot$ yr$^{-1}$. These properties suggest a possible connection between some high-redshift GRB host galaxies and high-$z$ massive sub-millimeter galaxies, which are characterized by disturbed morphologies and high metallicities. Our result provides additional evidence that the dispersion in the chemical enrichment of the Universe at high redshift is substantial, with the existence of very metal rich galaxies less than two billion years after the Big Bang.

\end{abstract}


\begin{figure*}
\includegraphics[width=170mm,angle=0]{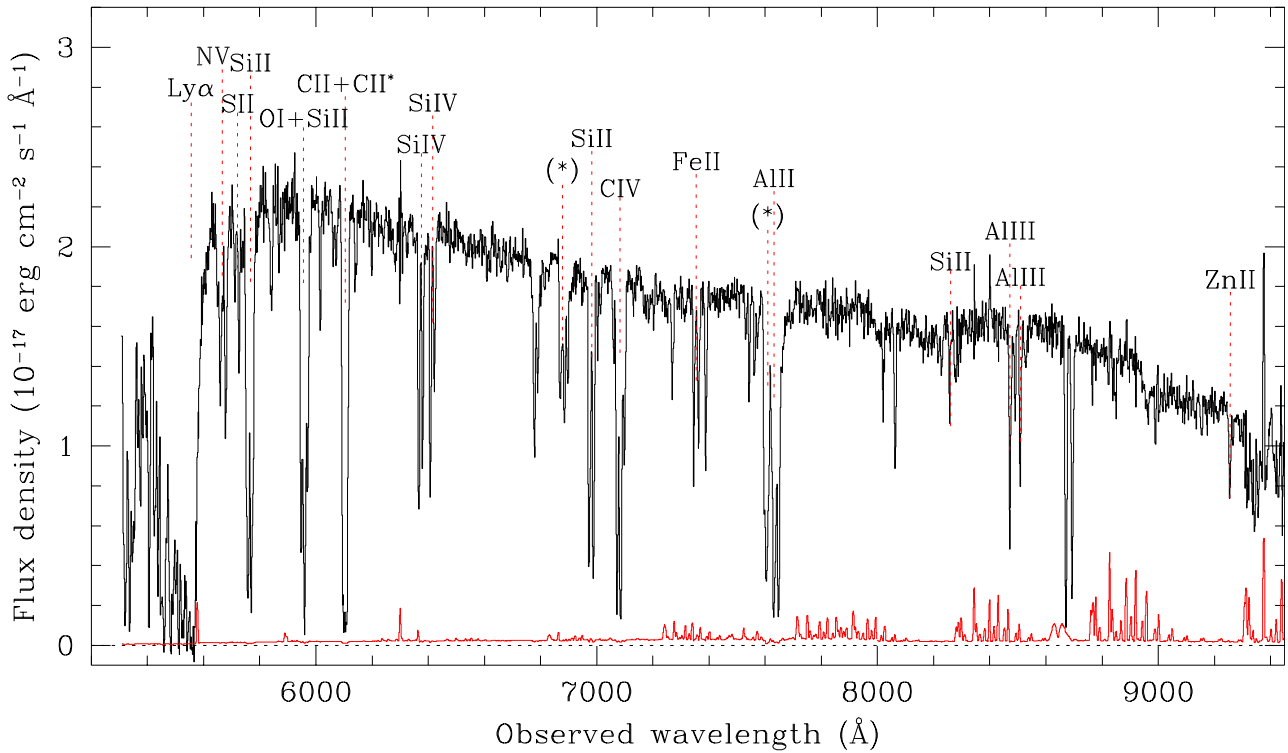}
\caption{The optical VLT/FORS2 spectrum of the afterglow of  \grb. Indicated are the main absorption lines associated with the two systems G0 and G1 at $z\sim 3.57$. Many other lines, associated with intervening systems are present, but not marked. Zn\,\textsc{ii} absorption is identified in G1. The red line at the bottom is the spectrum of the Earth atmosphere. The two stars indicate the telluric absorption.}
\label{ffors}
\end{figure*}

\section{Introduction}

The chemical enrichment history of galaxies is one of the key diagnostics of the evolutionary tracks of our Universe. The main element production factories since the Big Bang are stars, which transform light elements into heavier ones.   Understanding the cosmic chemical evolution requires detailed spectroscopic observations and theoretical models. For a long time, the high-redshift Universe was explored by analyzing absorption systems in galaxies distributed along the sight lines to high-redshift QSOs. Absorbers with high H\,\textsc{i} content ($N_{\rm H\,I} \gsim10^{20}$ \cm), called damped Lyman-alpha (DLA) systems (Prochaska \& Wolfe 1997), have predominantly neutral gas ($N_{\rm H} \simeq N_{\rm H\,I}$, temperature $T \lsim 1,000$ K). Their metallicity is generally low at high redshift, way below 1/10 solar at $z\sim4$, and approaches solar in the nearby Universe (Wolfe, Gawiser, \& Prochaska 2005;  Peroux et al.\ 2003). More recently, it has been possible to investigate the chemical enrichment in galaxies using emission lines, whose fluxes represent the integrated signal emitted by hot gas ($T \sim 10,000$ K) in H\,\textsc{ii} regions of star forming galaxies. About solar metallicity was detected in $2.1 < z < 2.4$ galaxies (Shapley et al. 2004). Similarly, the mass-metallicity relation detected in Lyman-break galaxies (LBGs) at $3.0 < z < 3.7$ indicates that the most massive objects ($M_\star \sim 10^{11}$ M$_\odot$) can be metal rich (Maiolino et al. 2008; Mannucci et al. 2009). The emerging picture is that the spread in metallicity is large at any redshift, certainly much larger than what one would expect based on DLAs in QSO spectra alone.

DLAs are regularly detected in spectra of gamma-ray burst (GRB) afterglows. Long GRBs are the most luminous explosions associated with the death of massive stars, and, as regularly done with QSOs, are used as background bright point-like sources for the investigation of the interstellar medium (ISM) in high-$z$ galaxies. The main differences to QSOs are that GRB afterglows can be brighter, are short lived and happen in star-forming regions. QSO-DLAs trace primarily metal-polluted galaxy haloes, where most of the volume of a galaxy is located and where dust extinction is less important (Fynbo et al.\ 2008). The first chemical analysis of absorption lines in the DLA detected in a $z \simeq 2$ GRB afterglow spectrum revealed a high (about solar) metallicity (Savaglio et al.\ 2003). This  neutral-ISM dominated GRB-DLA in a high-$z$ galaxy, the GRB host, opened a new view our understanding of the chemical enrichment of the Universe. In particular, the about solar abundance called into question the idea that low metal pollution in galaxies is the rule at high redshift. The typical GRB-DLA metallicity is indeed larger than in QSO-DLAs (Savaglio et al.\ 2003; Vreeswijk et al. 2004; Berger et al.\ 2006; Prochaska et al. 2007b; Fynbo et al. 2008), leading to the conclusion that  high-$z$ GRB hosts show properties more similar to UV bright star forming, LBG like, galaxies, than QSO-DLAs.

Here we report the discovery of two DLAs at \zr$\pm 0.0005$ (henceforth called G0) and \zb$\pm 0.0003$ (G1), separated in the velocity space by $\approx660$\,km s$^{-1}$, revealed in the rest-frame UV afterglow spectrum of \grb\ (Chornock  et al.\ 2009; Cenko et al.\ 2011; McBreen et al.\ 2010). The two absorbers (probing the Universe when it was 13\% of its present age) could belong to a single galaxy, to two distinct galaxies in close proximity and perhaps in the process of merging, or to two physically distant and unrelated galaxies. In the latter case, if the redshift difference is attributed to cosmic expansion alone, it would correspond to a separation of $\approx5.8$\,Mpc in comoving coordinates.

In this paper we analyze the spectrum of the \grb\ afterglow to probe the chemical enrichment of G0 and G1. Observations are described in \S2, the spectral analysis and heavy-element measurements in \S3 and \S4, respectively. The final discussion and conclusions are in \S5 and \S6, respectively. Throughout the paper we adopt an $H_0 = 70$ \kms, $\Omega_M = 0.3$, $\Omega_\Lambda = 0.7$ cosmology (Spergel et al.\ 2003).

\section{Observations}

\begin{figure}
\includegraphics[width=90mm,angle=0]{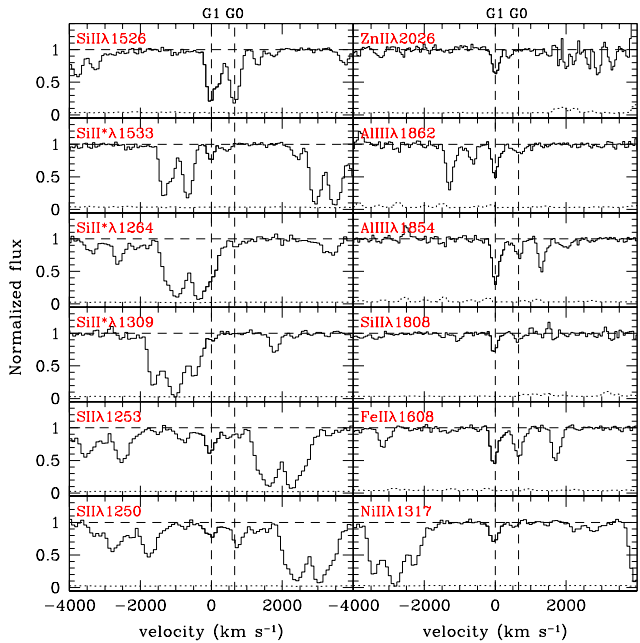}
\caption{The metal absorption lines associated with the neutral gas in G0 and G1, at $\Delta v = +660$ \kms\ and $\Delta v = 0$ \kms, respectively. The dotted line is the noise spectrum.}
\label{spectrum} 
\end{figure}

\grb\ was detected and localized by the Gamma-ray Burst Monitor (GBM) and the Large Area Telescope (LAT) onboard NASA's Fermi Gamma-ray Space Telescope (Michelson et al.\ 2008) on 23 March UT (Ohno et al.\ 2009). Observations by Swift's X-ray Telescope (XRT; Kennea et al.\ 2009) lead to the discovery of the afterglow and enabled longer-wavelength follow-up by  the Gamma-Ray burst Optical/Near-infrared Detector (GROND; Updike et al.\ 2009; McBreen et al.\ 2010), Palomar 60-inch Telescope (Cenko et al.\ 2009), and Gemini-S (Chornock et al. 2009). 

Optical spectroscopy of the  afterglow started on 2009 March 24.25 UT (1.2\,d  post-burst) using the FOcal Reducer and low dispersion Spectrograph 2 (FORS2) mounted  at the 8.2\,m ESO-VLT U1 telescope in Paranal, Chile, when the AB $R$-band magnitude of the afterglow was $\approx 18.6$ (McBreen et al.\ 2010). We obtained  $5\times900$\,s   integrations  using  the  600RI  grism
(5300\,\AA -- 8450\,\AA\ coverage) and $6\times900$\,s   integrations  using  the  600z  grism
(7500\,\AA -- 10,000\,\AA\ coverage) with a long slit of 1\farcs0 width. The bluer sequence started at 06:05 UT and ended at 08:13 UT, and was alternated with the redder sequence which started at 06:53 UT and ended at 09:01 UT the same night.
The average  seeing was $\approx0\farcs60$ resulting in  effective spectral  resolutions  of  $\approx4.6$\,\AA\  FWHM  at 6300\,\AA,  and $\approx4.8$\,\AA\  FWHM  at  8100\,\AA, or $\approx210$\,km\,s$^{-1}$ and $\approx150$\,km\,s$^{-1}$, respectively.

The data were reduced with standard IRAF routines, and  spectra  were extracted  using  an  optimal  (variance-weighted) method. Spectro-photometric  flux calibration  was  carried  out against  observations of the standard star LTT6248 obtained in the same night. Corrections for slit losses, due  to finite slit width and for Galactic foreground extinction, were applied. The individual exposures were combined and the final spectrum is shown in Figure~\ref{ffors}. In order to search for spectral time variability (in particular in the fine-structure transition Si\,\textsc{ii}$^*$), we also merged the first and second halves of the exposures individually and compared the strengths of the absorption lines. No variability was detected.

\begin{figure*}
\includegraphics[width=85mm,angle=0]{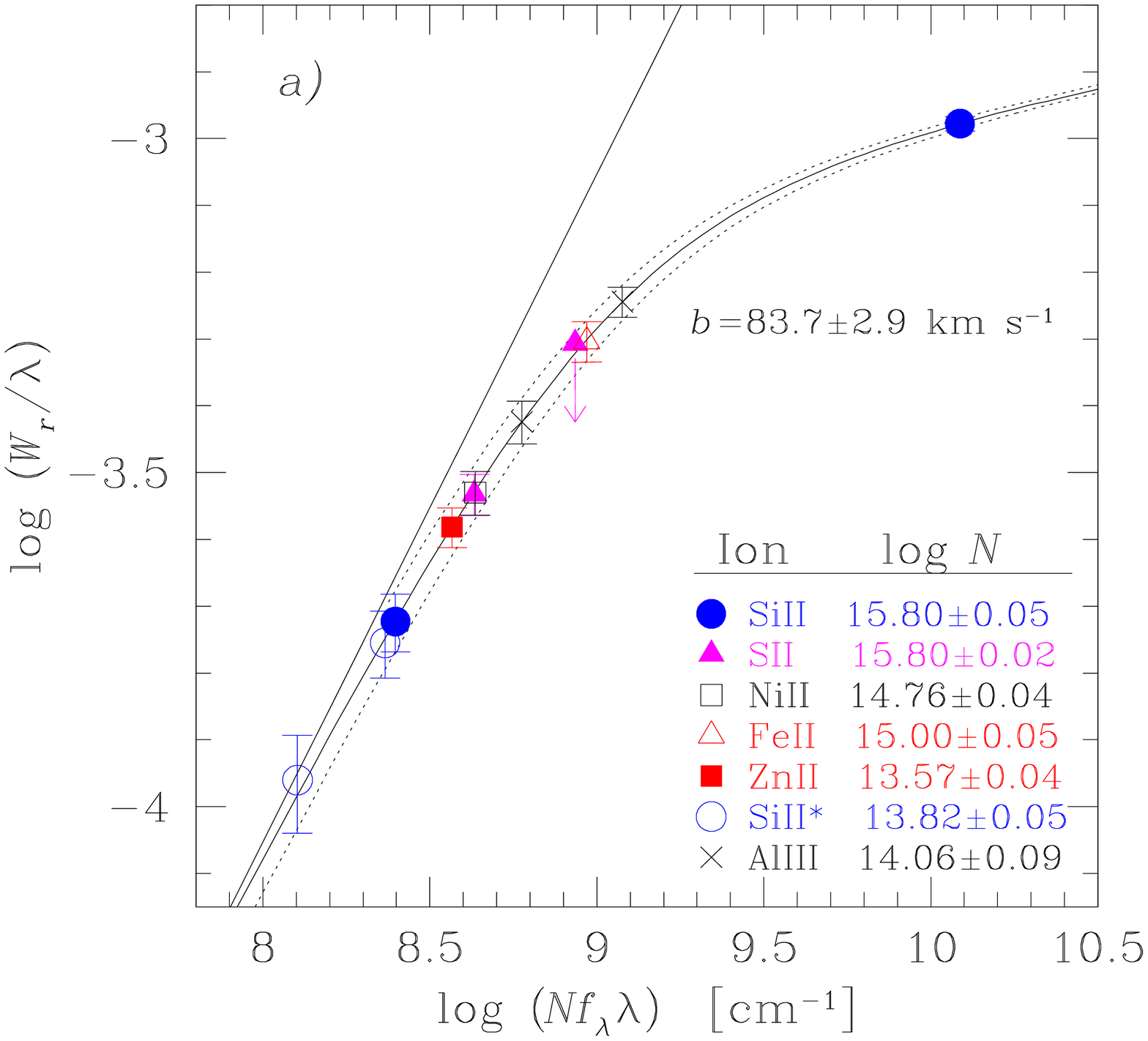}
\includegraphics[width=85mm,angle=0]{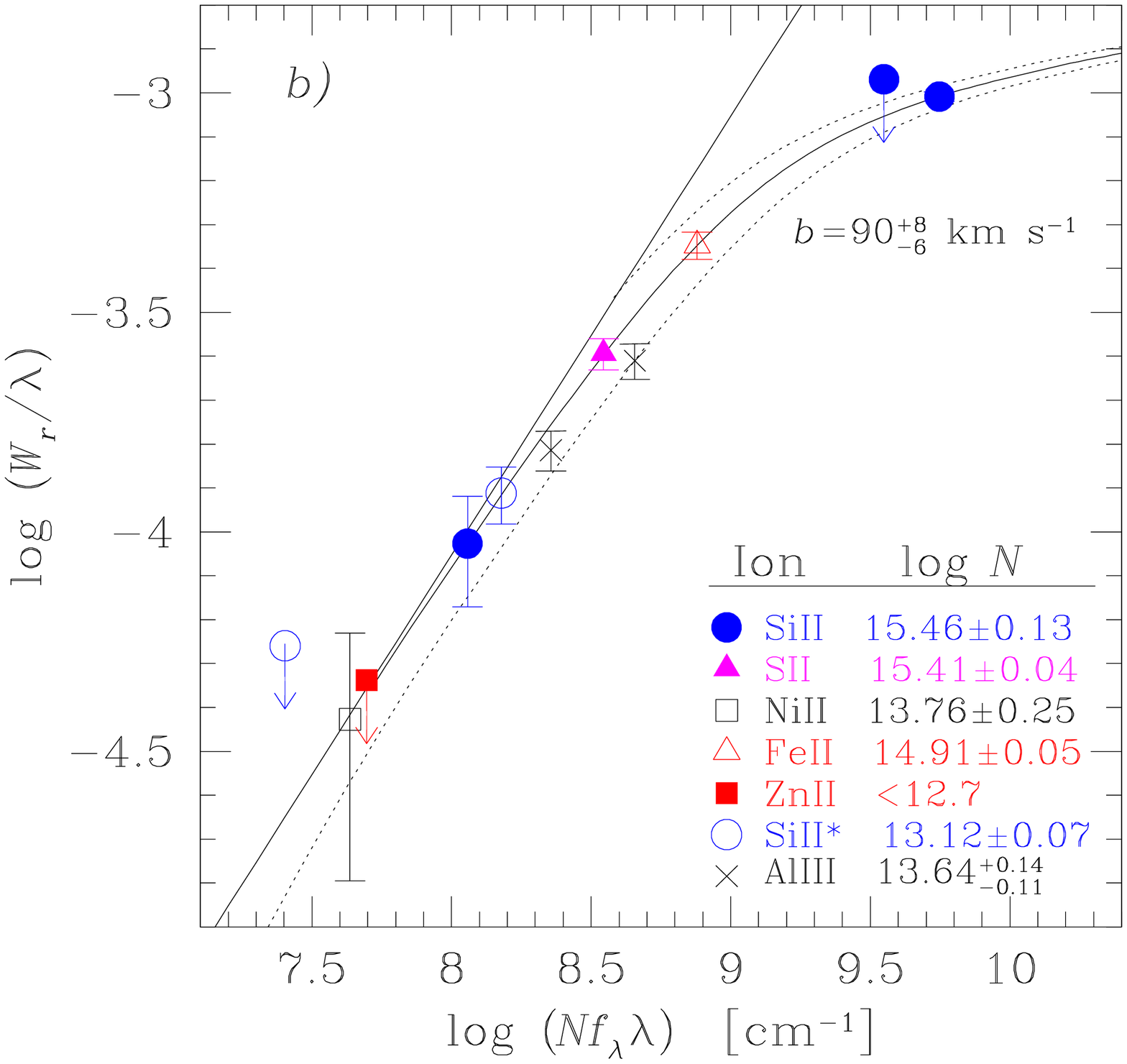}
\caption{Curve of growth analysis of low-ionization species in G1 (Figure \ref{curve}$a$) and G0 (Figure \ref{curve}$b$). In the $x$-axis label, $f_\lambda$ indicates the oscillator strength of the atomic transition. The best fit and 1$\sigma$ uncertainty are the continuous and dotted curves, respectively, while the straight line shows the linear approximation. The column density of the medium-ionization species Al\,\textsc{iii} for both systems is also included. For G1, the Doppler parameter $b\simeq84$ \kms\ has been derived from the  best fit of Si\,\textsc{ii}, which is consistent with the best-fit Doppler parameter of Si\,\textsc{ii}$^*$ and Al\,\textsc{iii}. For all other ions, this is assumed to be 84 \kms. Similarly for G0, the Doppler parameter $b\simeq90$ \kms\ has been derived from the best fit of Si\,\textsc{ii}. The best fit curve of growth for Al\,\textsc{iii} gives a very uncertain $b\simeq48$ \kms, but the best-fit column density is not very different if we force $b=90$  \kms, as for Si\,\textsc{ii}.} 
\label{curve} 
\end{figure*}

\section{Column densities of Absorption lines}

Details of some metal lines associated with the low-ionization species in G1 and G0 are shown in Figure~\ref{spectrum}. G0 is the reddest absorption system detected in the spectrum of \grb. However, given the possible strong interaction between the two, we cannot be sure that G0 is associated with the galaxy hosting \grb. Column densities (Tables~\ref{t1} and \ref{t2}) are estimated by using the apparent optical depth (AOD) method (Savage \& Sembach 1991) and curve of growth (COG) analysis (Spitzer 1978). These methods are routinely applied when high-resolution and signal-to-noise spectroscopy (for which the best fit of Voigt profiles is possible) is not available. The COG analysis (Figure~\ref{curve}) is robust for weak lines, when the linear approximation can be applied (e.g., rest-frame equivalent width $W_r < 0.2$ \AA\ and effective Doppler parameter $b>30$ km s$^{-1}$; Spitzer 1978). For stronger lines, the COG gives reliable results if several lines with different oscillator strengths of the same ion are detected. This idea is supported, for instance, by the spectral analysis of the afterglow of GRB\,081008 by D'Elia et al.\ (2011), who have shown that column densities derived from a low resolution spectrum with the COG are consistent with those derived from Voigt profile fitting of a high resolution spectrum. The AOD method is used for lines with moderate to low equivalent widths (EWs), provided that the effective Doppler parameter is not very low ($ b \gsim 10$ \kms; Prochaska et al.\ 2007a). COG values (adopted for our final abundance determination) are generally consistent with AOD ones (Tables~\ref{t1} and \ref{t2}). The AOD gives at times slightly lower values, indicating that residual saturation is not always taken into account. 
Prochaska (2006) has shown that column densities of metal lines derived with the COG can be underestimated.  If this problem affected our measurements, the true metallicities would be even larger than those reported here.

\begin{figure}
\includegraphics[width=62mm,angle=-90]{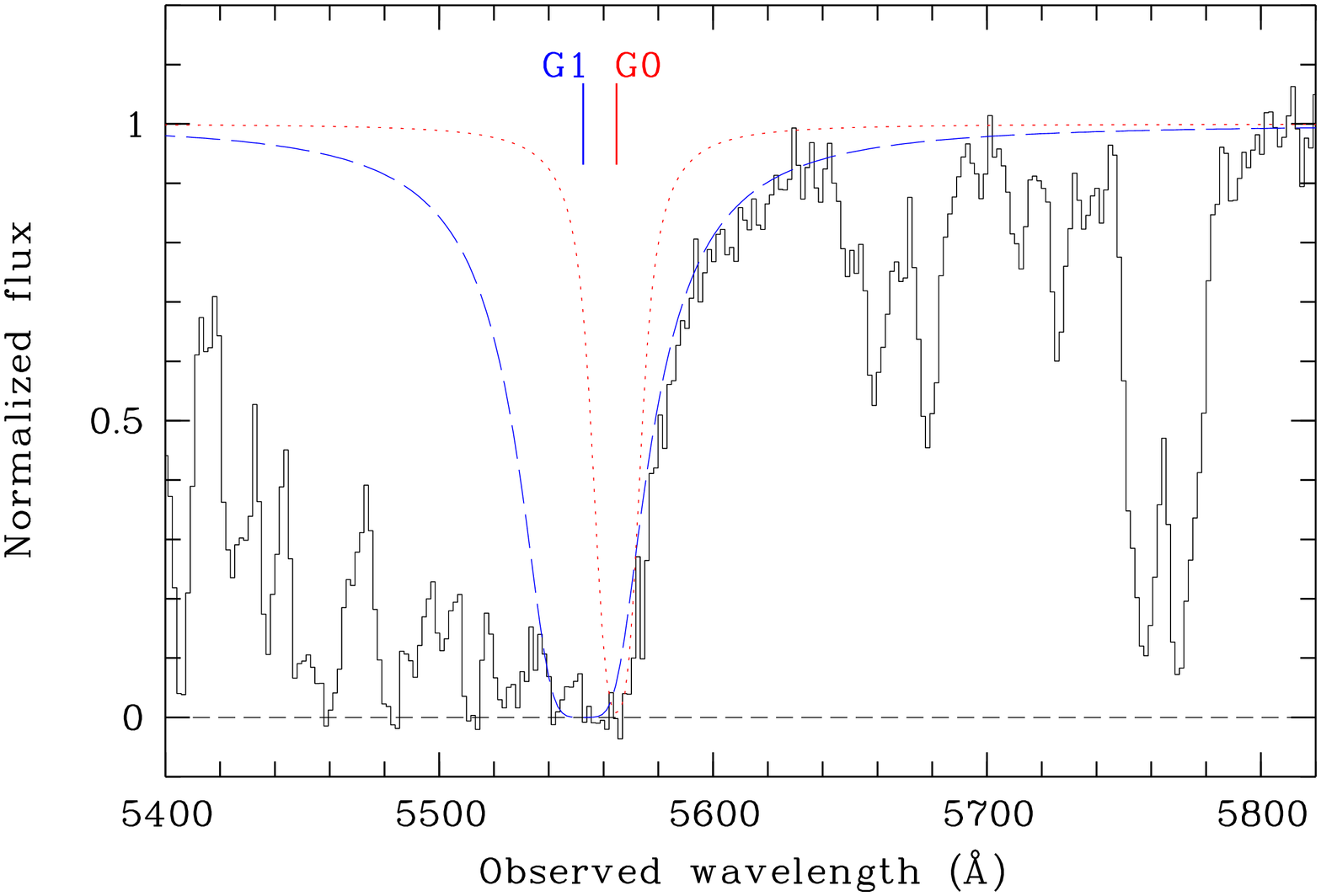}
\includegraphics[width=62mm,angle=-90]{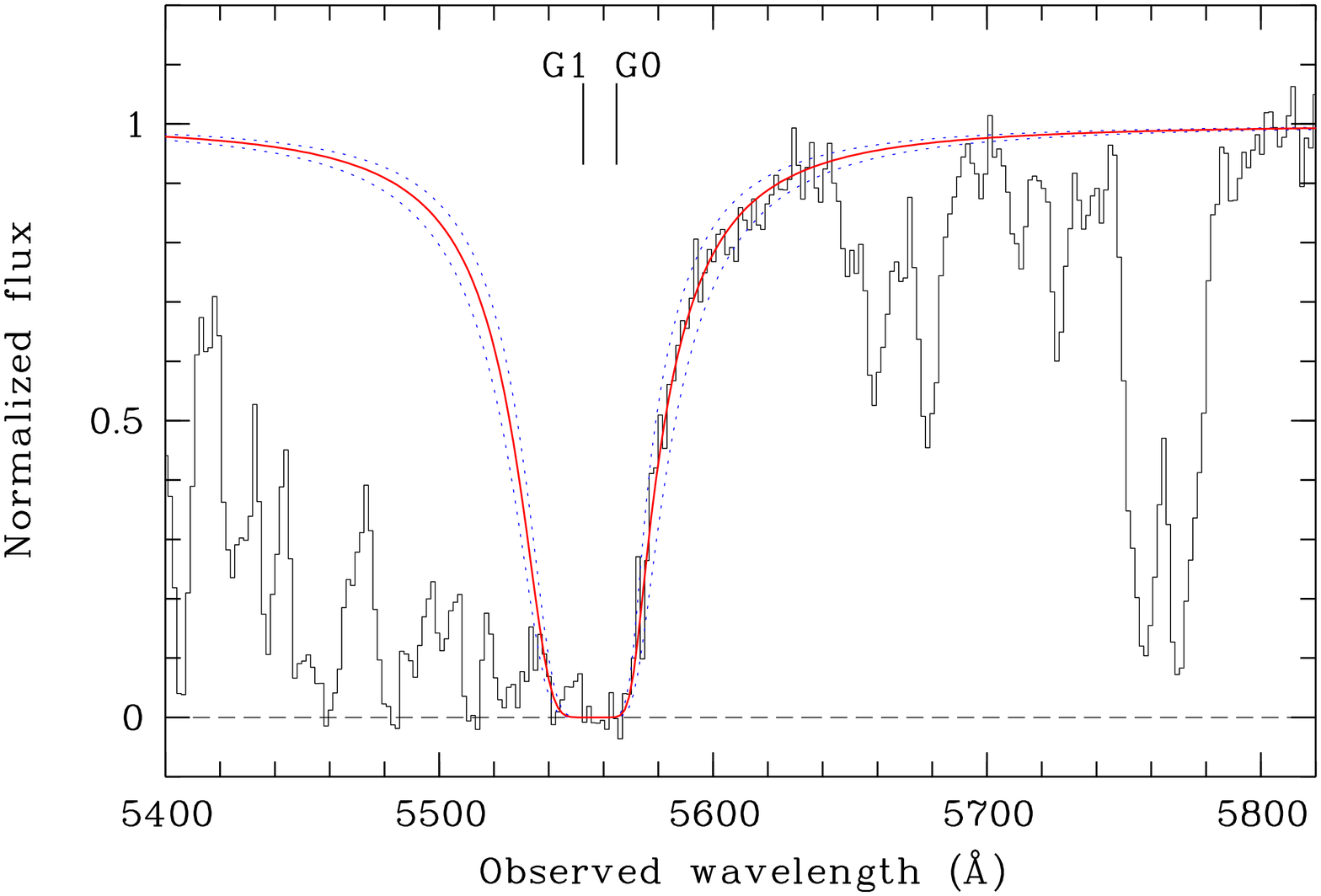}
\caption{The two panels show the FORS2 spectrum of \grb\ around the Ly$\alpha$ region. The two systems at \zb\ (G1) and \zr\ (G0) are indicated by the two vertical tick marks. {\it Upper panel:} the dashed-blue and red-dotted lines indicate the individual Ly$\alpha$ profiles with $N_{\rm H\,I,G1}=10^{20.72\pm0.09}$ \cm\ and $N_{\rm H\,I,G0}=10^{19.62\pm0.33}$ \cm, respectively. From these, we derive ${\rm [Zn/H]} = +0.29\pm0.10$ and ${\rm [S/H]} = +0.67\pm0.34$. {\it Lower panel:} the red solid line is the total best-fit profile. The two dotted lines represent two distinct solutions contained in the 90\% confidence interval. These are $N_{\rm H\,I, G1}=10^{20.81}$ \cm\ and $N_{\rm H\,I,G0}=10^{19.96}$ \cm, and $N_{\rm H\,I, G1} =10^{20.63}$ \cm\ and  $N_{\rm H\,I,G0}=10^{19.48}$ \cm.}
\label{flya}
\end{figure}

The H\,\textsc{i} column densities are derived from the Lyman-$\alpha$ spectral absorptions (Figure~\ref{flya}). These are strongly blended and show the characteristic damping wing typical for column densities $ > 10^{19}$ \cm.  The Voigt-profile best fit gives $N_{\rm H\,I} = 10^{20.72\pm0.09}$ \cm\ and $N_{\rm H\,I} = 10^{19.62\pm0.33}$ \cm\ for G1 and G0, respectively (at fixed redshifts, as derived for the metal systems). As indicated by Figure~\ref{flya}, there is virtually no  lower bound to $N_{\rm H\,I}$ in G0. However, the presence of strong absorption lines associated with heavy-element low-ionization transitions is an indication that  $N_{\rm H\,I}$ cannot be very low. The upper limit for the metallicity in G1 is set by the requirement that here the metallicity cannot be higher than 10 times solar.

\begin{figure}
\includegraphics[width=90mm,angle=0]{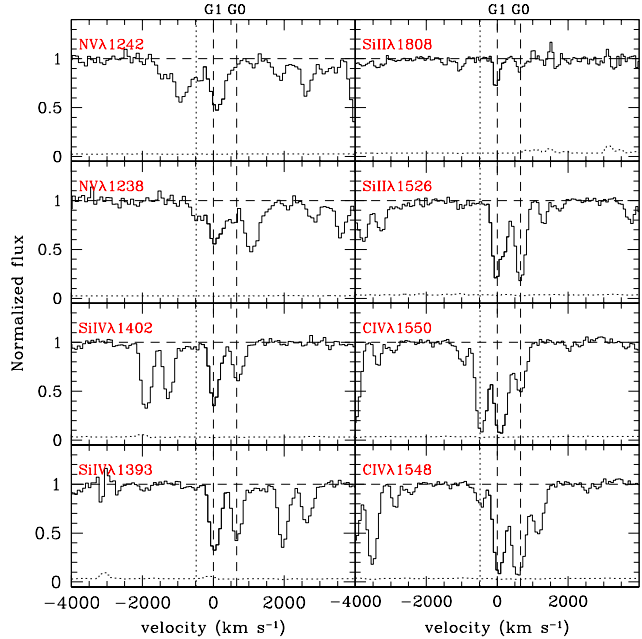}
\caption{High-ionization absorptions in \grb. The two strongest absorbers are marked by the two vertical dashed lines. A possible third weak absorber, shown by C\,\textsc{iv}$\lambda1548$ and N\,\textsc{v}$\lambda1239$ features, at $z=3.5600$ ($\Delta v \simeq 500$ \kms\ blueward of G1) is also marked by the vertical dotted line. The horizontal dotted line is the noise spectrum.}
\label{metal} 
\end{figure}

In Figure~\ref{metal}, we also display the velocity profiles of high-ionization lines associated with G0 and G1 (e.g., N\,\textsc{v}, Si\,\textsc{iv}, and C\,\textsc{iv}). Together with G0 and G1, we mark a possible weak absorption system $\sim 500$ km s$^{-1}$ blueward of G1, with the tentative identification of N\,\textsc{v}$\lambda1239$ and C\,\textsc{iv}$\lambda1548$ absorptions. The VLT spectra also reveal a wealth of strong metal absorption lines associated with absorbers in galaxies distributed along the GRB sight line, at redshift $z\gsim1$. These are at $z=1.697$, 2.101, 2.448, and 3.379, and not considered further in this study. 

\begin{table*}
 \centering
 \begin{minipage}{140mm}
\caption[t1]{Element abundances of G1 at \zb\ ($N_{\rm H\,I} = 10^{20.72\pm0.09}$ \cm).}
\begin{tabular}{lccccc}
\hline\hline&&&&&\\[-5pt]
Ion  & $\lambda_{\rm obs}$ & $W_r$  & \multicolumn{2}{c}{$\log N_{\rm X}$ [\cm]} & [X/H] \\
\cline{4-5}\\[-7pt] 
&(\AA) &(\AA) & AOD & COG (adopted) & \\
[3pt]\hline&&&&&\\[-5pt] 
Zn\,\textsc{ii}$\lambda2026$$^{a,b}$ & 9254.0 & $0.55\pm0.04$ & $13.58\pm0.03$ & $13.57\pm0.04$ & $+0.29\pm0.10$  \\
S\,\textsc{ii}$\lambda1250$ 	& 5711.8 & $0.37\pm0.03$ 	& $15.72\pm0.03$ & $15.80\pm0.02$ & $-0.04\pm0.09$ \\
S\,\textsc{ii}$\lambda1253$$^c$ & 5726.5	& $<0.6$ 	& . . .	& . . . &  . . .  \\
O\,\textsc{i}$\lambda 1302$ & 5947.4	& $\sim1.5$	& $>15.5$		& $>16.6$ & $>-0.8$  \\
Ni\,\textsc{ii}$\lambda1317$ & 6016.1	& $0.39\pm0.03$	&  $14.69\pm0.03$ & $14.76\pm0.04$ & $-0.18\pm0.10$ \\
Ni\,\textsc{ii}$\lambda1370$ & 6257.8	& $<0.11$ & . . . &  $<14.1$   & . . .   \\
Si\,\textsc{ii}$^*\lambda1309$ & 5979.8 & $0.14\pm0.02$ 	& $13.83\pm0.08$ & . . . & . . .  \\
Si\,\textsc{ii}$^*\lambda1533$ & 7003.6 & $0.27\pm0.03$	& $13.80\pm0.06$ & $13.82\pm0.05$ & . . .   \\
Si\,\textsc{ii}$\lambda1304$$^d$ & 5957.4	& $<2.2$	& . . . 	& . . .                          & . . .   \\
Si\,\textsc{ii}$\lambda1526$ & 6972.9	& $1.61\pm0.04$	& . . . 	& . . .       & . . .  \\
Si\,\textsc{ii}$\lambda1808$ & 8257.7	& $0.34\pm0.03$	& $15.79\pm0.04$& $15.80\pm0.05$ & $-0.43\pm0.10$  \\
Fe\,\textsc{ii}$\lambda1608$  & 7346.3 & $0.80\pm0.06$	&  $14.89\pm0.03$& $15.00\pm0.05$ & $-1.22\pm0.10$ \\
Al\,\textsc{ii}$\lambda1670$	& 7631.0 & $<2.7$			& . . .	& $<16$                   & $<+0.8$  \\
Al\,\textsc{iii}$\lambda1854$ & 8471.0	& $1.06\pm0.05$	&  . . .& . . . & . . .    \\
Al\,\textsc{iii}$\lambda1862$ & 8507.9	& $0.70\pm0.05$	&  $14.01\pm0.04$ & $14.06\pm0.09$  & . . .   \\
[-1pt]\hline
\end{tabular}
\\
$^a$An estimated small contamination by Mg\,\textsc{i}$\lambda2026$ ($W_r \simeq 0.031$ \AA) is subtracted from the measured EW. Cr\,\textsc{ii}$\lambda2026$ is very weak and can be neglected (see text).\\
$^b$Zn\,\textsc{ii}$\lambda2062$ is not detected because of blending with strong telluric lines.\\
$^c$Contaminated by S\,\textsc{ii}$\lambda1250$ of G0.\\
$^d$Contaminated by O\,\textsc{i}$\lambda1302$ of G0.\\
\label{t1}
\end{minipage}
\end{table*}

\begin{table*}
 \centering
 \begin{minipage}{140mm}
\caption[t1]{Element abundances of G0 at \zr\ ($N_{\rm H\,I} = 10^{19.62\pm0.33}$ \cm).}
\begin{tabular}{lccccc}
\hline\hline&&&&&\\[-5pt]
Ion  & $\lambda_{\rm obs}$ & $W_r$  & \multicolumn{2}{c}{$\log N_{\rm X}$ [\cm]} & [X/H] \\
\cline{4-5}\\[-7pt] 
&(\AA) &(\AA) & AOD & COG (adopted) & \\
[3pt]\hline&&&&&\\[-5pt] 
Zn\,\textsc{ii}$\lambda2026$$^a$ & 9274.4 & $< 0.09$	& . . .	 & $<12.7$ & $<+0.6$  \\
S\,\textsc{ii}$\lambda1250$$^b$  & 5724.4	& $<0.6$ 			& . . . & $<16.0$  & . . .   \\
S\,\textsc{ii}$\lambda1253$ & 5739.2	& $0.32\pm0.03$ 	& $15.35\pm0.04$ & $15.41\pm0.04$ & $+0.67\pm0.34$   \\
Ni\,\textsc{ii}$\lambda1317$ & 6029.4	& $0.05\pm0.03$	&  $13.77\pm0.27$ & $13.76\pm0.25$ & $-0.09\pm0.42$   \\
Ni\,\textsc{ii}$\lambda1370$ & 6271.6	& $<0.2$		&  . . . & $<14.3$  & . . .    \\
Si\,\textsc{ii}$^*\lambda1264$ & 5789.2 & $0.15\pm0.02$ 	& $13.10\pm0.07$ & $13.12\pm0.07$ & . . .  \\
Si\,\textsc{ii}$^*\lambda1309$ & 5993.1 & $<0.07$ 	& . . . & $<13.5$ & . . .   \\
Si\,\textsc{ii}$^*\lambda1533$  & 7019.1 & $<0.10$	& . . . & $<13.3$ & . . .  \\
Si\,\textsc{ii}$\lambda1304$ & 5970.6 	& $<1.4$	& . . . 	& . . .                          & . . .   \\
Si\,\textsc{ii}$\lambda1526$ & 6988.3 & $1.50\pm0.04$	& . . . 	& . . .                          & . . .  \\
Si\,\textsc{ii}$\lambda1808$  & 8276.0	& $0.17\pm0.05$	& $15.45\pm0.13$& $15.46\pm0.13$ & $+0.33\pm0.36$  \\
Fe\,\textsc{ii}$\lambda1608$ & 7362.5 & $0.72\pm0.05$	&  $14.82\pm0.03$ & $14.91\pm0.05$ & $-0.21\pm0.34$   \\
Al\,\textsc{ii}$\lambda1670$ & 7647.9	& $<2.4$			& . . .		& $<15.4$                   & $<+1.3$  \\
Al\,\textsc{iii}$\lambda1854$ & 8489.8	& $0.45\pm0.04$	& . . . 			& $13.64_{-0.11}^{+0.14}$ & . . .  \\
Al\,\textsc{iii}$\lambda1862$	 & 8526.7 & $0.29\pm0.03$	&  $13.55\pm0.05$	& . . .  & . . .  \\
[-1pt]\hline
\end{tabular}
\\
$^a$An upper limit to Zn\,\textsc{ii}$\lambda2062$ is not estimated because of blending with strong telluric lines.\\
$^b$Contaminated by S\,\textsc{ii}$\lambda1253$ of G1.\\
\label{t2}
\end{minipage}
\end{table*}

\section{Heavy-element abundances}

In DLAs, the ionization correction can generally be neglected (Dessauges-Zavadsky et al.\ 2003; Jenkins 2009), and ions with ionization potentials just above the ionization energy of hydrogen dominate all other states (e.g., $N_{\rm Fe\,II} \sim N_{\rm Fe}$, $N_{\rm Zn\,II} \sim N_{\rm Zn}$, $N_{\rm S\,II} \sim N_{\rm S}$). The chemical state of the gas is thus derived using the simple approximation ${\rm [X/H]} =  \log (N_{\rm X\,II} / N_{\rm H\,I}) - \log (N_{\rm X} /N_{\rm H})_\odot$ (Pettini et al.\ 1997). Among the many detected lines, the interesting Zn\,\textsc{ii}$\lambda2026$ line in G1 at \zb\ is detected with high confidence (Figure~\ref{spectrum}) thanks to the high signal-to-noise in the spectrum.

The H\,\textsc{i} column density of G0 is relatively low ($N_{\rm H\,I} = 10^{19.62}$ \cm), for which the ionization correction could be necessary. The ionization correction is generally smaller than 0.2 dex in DLAs with $19. 5 < \log (N_{\rm H\,I}/{\rm cm^{-2}}) <20.3$, and smaller for larger H\,\textsc{i} columns (Dessauges-Zavadsky et al.\ 2003; P{\'e}roux et al.\ 2007). The low ionization in our systems is supported by large column densities of low-ionization metal lines: Kanekar et al.\ (2009) found an anti-correlation between the gas temperature and metallicity of DLAs. The relatively low column density of Al\,\textsc{iii} (Table~\ref{t2}) additionally indicates low ionization (Meiring et al.\ 2008). Although we can only infer an upper limit for Al\,\textsc{ii}, the expected value can be estimated from  other ions with similar ionization potential, from which we conclude that  $N_{\rm Al\,II} / N_{\rm Al\,III} > 10$. Some heavy elements (e.g., iron or nickel) can be partly locked up in dust grains, and measured column densities (which are for the gas phase only) can be very different from the total column density. Sulfur and zinc are mainly in the gas phase, so that uncertain dust depletion corrections are less critical.

\begin{figure}
\includegraphics[width=86mm,angle=0]{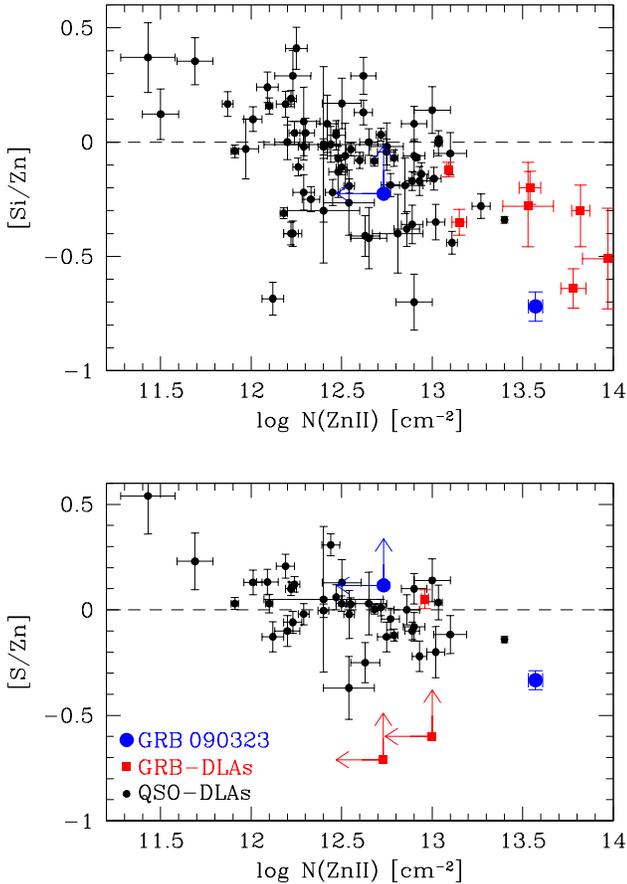}
\caption{Silicon-to-zinc and sulfur-to-zinc relative abundances (upper and lower panel, respectively) as a function of Zn\,\textsc{ii} column density. Limits for G0 (with $N_{\rm Zn\,II}<10^{12.7}$ \cm) and values for G1 (with $N_{\rm Zn\,II}=10^{13.6}$ \cm) are the blue filled circles. Other GRB-DLAs are represented as red squares, QSO-DLAs are black dots. Dashed  lines indicate solar relative abundances.}
\label{szn} 
\end{figure}

Figure~\ref{szn} shows the abundances of Si and S relative to Zn as a function of Zn\,\textsc{ii}  column density for G0 and G1, other GRB-DLAs and QSO-DLAs. For G1 we derive [S/Zn] $=-0.33\pm0.04$ and  [Si/Zn] $=-0.72\pm0.06$. Anomalous relative abundances have been found before in high-$z$ absorption line systems. In QSO-DLAs, significant [S/Zn] $<0$ is observed for about half of all systems where these two elements are detected. Cooke et al.\ (2011) report on a QSO-DLA with [S/C] $< -1.51$. The metal column densities, derived from a composite spectrum of 60 GRB afterglow spectra, indicate [S/Zn] $= -0.94$ (Christensen et al.\ 2011). Nissen et al.\ (2004) report low S-to-Zn ratios in QSO-DLAs with respect to Galactic stars, from which a different star-formation history is hypothesized. Nissen et al.\ (2007) were able to lower [S/Zn] values in Galactic stars by using non-LTE modeling. Vladilo et al.\ (2011) suggest that the observed deficit of S is indicative of a larger dust depletion with respect to Zn in dust rich sight lines. These anomalies are not totally understood, another contributing factor might be a theoretically very uncertain element yields in early generations of massive and low-metallicity stars.

The Zn\,\textsc{ii}$\lambda2026$ line in G1 is contaminated by Mg\,\textsc{i} and Cr\,\textsc{ii}. ÊCr\,\textsc{ii}$\lambda2026$ is weak and can be neglected. Mg\,\textsc{i}$\lambda2026$ is rarely detected at high spectral resolution and columns are in the interval $N_{\rm Mg\,I} = 10^{12.3-13.2}$ \cm.Ê To estimate the Mg\,\textsc{i}$\lambda2026$ contamination, we considered EWs measured in the GRB  afterglow composite spectrum of Christensen et al.\ (2011). The EWs of metal lines in G1 and the composite are similar, in particular for weaker lines ($W_r<1.1$ \AA).  From this, we assumed a possible $W_r=Ê0.8\pm0.3$ \AA\ for MgI$\lambda 2852$ in G1 (this is not covered by our spectrum). The error is given by the dispersion of the EWs for G1 and GRB composite  around the equality relation (80\% of the lines). Thus, from MgI$\lambda 2852$ and the COG (we assumed $b=84$ \kms, as for other lines), we estimated $W_r= 0.031^{+0.017}_{-0.014}$ \AA\ for MgI$\lambda 2026$,  which we subtracted from the Zn\,\textsc{ii}$\lambda2026$ EW (Table~\ref{t1}). The Cr\,\textsc{ii}$\lambda2056$ line is in a noisy part of the spectrum (sky emission contamination). We set a conservative limit $N_{\rm Cr\,II} < 10^{14.4}$ \cm, which gives [Cr/Zn] $< -0.3$, explainable by a possible depletion of Cr into dust grains.

A Wolf-Rayet (WR) star origin was advocated to explain the presence of double absorption features in GRB spectra. This could also explain the presence of the two strong nearby absorbers G0 and G1 in \grb. The GRB would take place within the dense, fast-moving shell of low-ionized gas. This scenario is disfavored (Chen et al.\ 2007), because the GRB strongly ionizes the circumburst environment out to a kpc distance scale (Vreeswijk et al.\ 2007; D'Elia et al.\ 2009; Ledoux et al.\ 2009), and thus large column densities of low ionized gas in the wind, as indicated by our observations, are unlikely. Local WR stars indeed do not show strong low-ionization absorption lines. Furthermore, even if hydrogen remained neutral in the blast, and 20 $M_\odot$ of hydrogen were ejected spherically to a distance of $\sim 10$ pc, we would expect  a H\,\textsc{i} column density of $6\times10^{18}$ cm$^{-2}$, which is well below the measured value.

The association of G0 and G1 with a single star forming region and its wind is also unlikely. We considered a standard SN remnant scenario and, under the assumption of a shell thickness of $\sim 10$\% of the shell radius, compute the swept-up mass. We consider ISM densities over a wide range, from 1 cm$^{-3}$ to 1500 cm$^{-3}$, and compute the shell radius for the time when the shell velocity differential reaches the observed 660 \kms. The resulting swept-up mass is of the order of 1000 M$_\odot$. A single SN cannot enrich this amount of swept-up ISM mass to a metallicity a few times solar. Even a situation of triggered star formation, with dozens SNe during a small time interval cannot change this picture, since each SN has to sweep up its 800-1000 M$_\odot$ to slow down to the observed velocity spread of 660 \kms.

\subsection{Fine-structure absorption in G0 and G1}

\begin{figure*}
\includegraphics[width=160mm,angle=0]{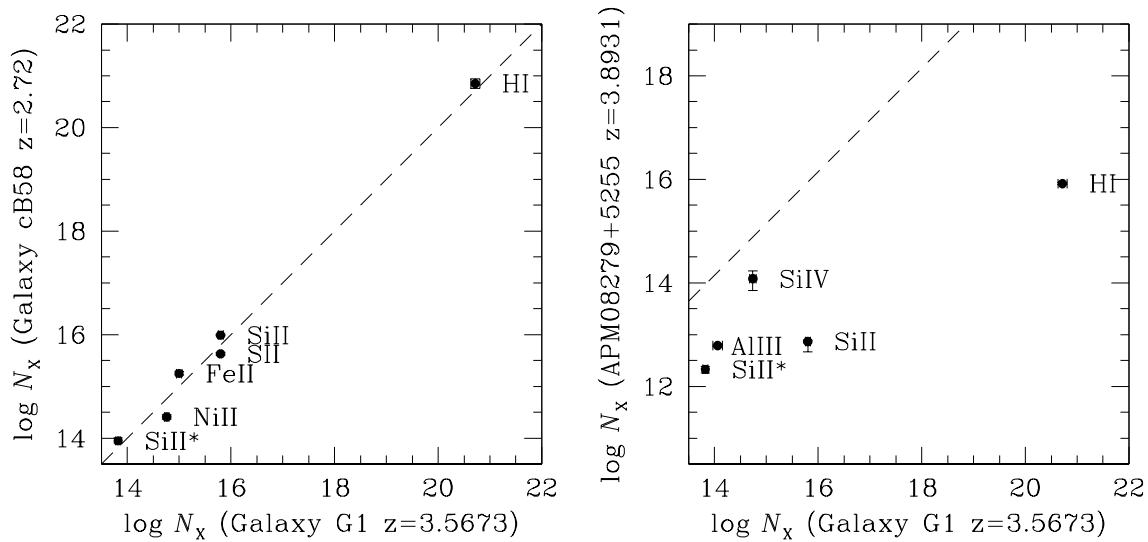}
\caption{The comparison between the column densities of G1 and cB58 (left panel), and G1 and QSO~APM08279+5255 (right panel). The straight lines on the left and right panels indicate the $x  \equiv y$ relations. This shows the similarities between the two DLAs in G1 and cB58 and the strong difference between G1 and narrow absorption lines close to QSO regions. Column densities for cB58 are recalculated by Savaglio (2006) and consistent with those reported by Pettini et al. (2002).}
\label{cb58}
\end{figure*}

Absorption lines associated with excited and metastable transitions are regularly reported in GRB afterglow spectra (Ledoux et al.\ 2009). The first excited transitions $J_{3/2}$ associated with the fine-structure species Si\,\textsc{ii}$^*$ are detected both in G0 and G1 (Tables~\ref{t1} and \ref{t2}, and Figure~\ref{spectrum}) with column densities $N_{\rm Si\,II^*} = 10^{13.12\pm0.07}$ \cm\ and  $10^{13.82\pm0.05}$ \cm, respectively. No other excited or metastable transitions are observed (some lines are likely blended with strong ground-level lines). In QSO-DLAs, C\,\textsc{ii}$^*$, but not  Si\,\textsc{ii}$^*$, is regularly observed (Wolfe et al.\ 2003). 
Figure~\ref{cb58} (left panel) shows the column densities of several ions (including Si\,\textsc{ii}$^*$) in G1 and the star-forming galaxy cB58 at $z=2.72$ (Pettini et al.\ 2002). The similarity between the column densities in G1 and cB58 suggests that the physical conditions of the neutral gas in G1 are not dissimilar from those of high-$z$ star-forming galaxies. In addition, Pettini et al.\ (2002) measured a relatively high metal enrichment: [S/H] $=-0.36\pm0.09$. The right panel of Figure~\ref{cb58} shows the comparison between the ionized gas in G1 and close to the QSO APM 08279+5255 at $z=3.9$ (Srianand \& Petitjean 2000). The Si\,\textsc{ii}$^*$-to-Si\,\textsc{ii} ratio in G0 and G1 is small (0.005 and 0.01, respectively), much lower than in APM 08279+5255, where this is $\sim0.3$, and indicating a high ionization of the gas near the massive black hole. The typical Si\,\textsc{ii}$^*$-to-Si\,\textsc{ii} ratio measured in GRB-DLAs is $\gsim 0.01$. Following the  modeling of Srianand \& Petitjean (2000), we conclude that the absorbing gas in G0, G1 and cB58 has a low density ($<100$ cm$^{-3}$) and/or a low temperature ($T<10^3$ K).

\subsection{Dust depletion and extinction}

McBreen et al.\ (2010) and Cenko et al.\ (2011) derived the optical dust extinction in the sight line of \grb\  by using the SED of the afterglow. This is relatively flat redward of the Ly$\alpha$, indicating a moderate $A_V = 0.1-0.2$. We alternatively estimated $A_V$ from the metal column densities of several elements with different refractory properties. These give information on the depletion of metals onto dust grains and thus the expected dust extinction. 

Detected absorption lines indicate that the contribution to extinction of G0 can be neglected (column densities  are weaker than in G1). To estimate the possible extinction in G1, we use the method described in Savaglio \& Fall (2004). As a reference value, we take the rate of extinction per column of neutral gas derived for the Milky Way  (Bohlin et al. 1978):

\begin{equation}\label{eq1}
A_V^{\rm MW}  = 0.53 \times \frac{N_{\rm H\,I}}{10^{21} {\rm cm^{-2}}} ~.
\end{equation} 

\noindent
This choice of Milky Way extinction rate is supported by a relatively low value of the iron-to-zinc ratio [Fe/Zn] $= -1.5$ in G1, not too dissimilar to the typical ratio in the ISM of the Milky Way (Savage \& Sembach 1991). Eq.\,\ref{eq1} can also be expressed in terms of the column of metals $N^{\rm met}$, by assuming that the rate of extinction in the optical per column of dust $A_V/N^{\rm dust}$ is proportional to $A_V/N^{\rm met}$ (Savaglio \& Fall 2004), or:

\begin{equation}\label{eq2}
A_V^{\rm MW}  = 0.53 \times \frac{N^{\rm met}}{N_{\odot, 21}^{\rm met}}
\end{equation}

\noindent
where $N_{\odot, 21}^{\rm met}$ is the metal column in a gas cloud with $N_{\rm H\,I} = 10^{21}$ \cm\ and solar metallicity. Eq.\,\ref{eq2} gives $A_V^{\rm MW} = 0.53$ for a gas cloud with zinc column density $N_{\rm Zn} = 10^{13.56}$ \cm. Our Zn\,\textsc{ii} column density for G1 is $N_{\rm Zn\,II} = 10^{13.57}$ \cm, which means that  $A_V \approx 0.5$ is expected. This optical extinction is higher than that measured from the SED ($A_V = 0.1-0.2$), and the typical $A_V$ derived from GRB afterglow SEDs (Zafar et al.\ 2011). The relatively flat \grb\ afterglow SED cannot be explained by a possible large column of dust (see also a similar conclusion in Savaglio \& Fall 2004), unless a sufficiently grey extinction is assumed (Stratta et al.\ 2004). However, a grey extinction is not confirmed by the mean dust extinction law derived from a sample of 17 GRB afterglow sight lines with X-ray-to-optical SEDs (Schady et al.\ 2011). 

\section{Discussion}

In G1 we found the most distant Zn\,\textsc{ii} absorption ever detected up to now, being followed by a DLA at $z=3.39$ in the spectrum of QSO~0000-2620 (Molaro et al.\ 2000). Zinc is widely considered to be the most reliable indicator of the metal abundance of DLAs in high-redshift galaxies (Pettini et al.\ 1997), and preferred over more abundant elements (e.g.\ iron, silicon or oxygen) because it is much less depleted to the dust component of the ISM, and because the spectral analysis is less affected by saturation. The column density $N_{\rm Zn\,II} = 10^{13.57\pm0.04}$ \cm\ in G1 is also among the highest ever inferred, with other systems of comparable density all residing at $z<2.9$. The large column densities in G1 would indicate the galaxy to be massive and/or metal rich. At this high redshift, this would make G1 a rare object.

\begin{figure}
\includegraphics[width=85mm,angle=0]{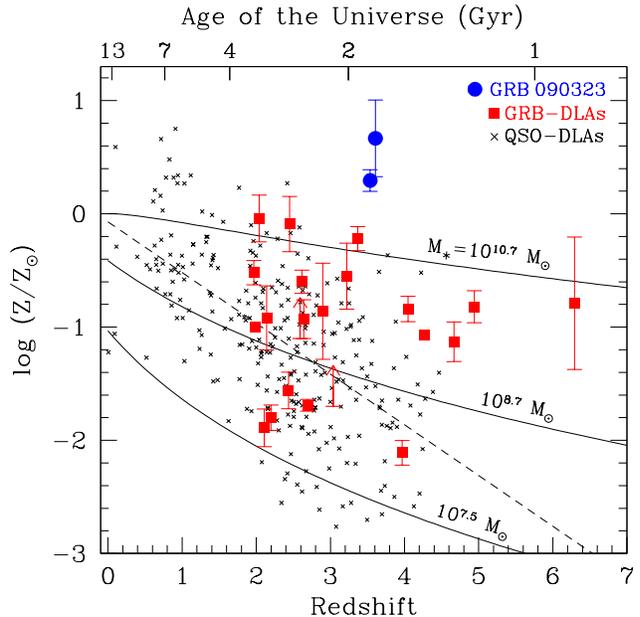}
\caption{Metallicities as a function of redshift of cold ISM in galaxies. G1 and G0 metallicities in \grb\ (left and right blue dots, respectively, artificially separated by $\Delta z = 0.06$ for clarity) are given by ${\rm [Zn/H]} = +0.29\pm0.10$ and ${\rm [S/H]} = +0.67\pm0.34$, respectively. Metallicities of other GRB-DLAs and QSO-DLAs are red squares and black crosses, respectively. GRB-DLA metallicities are collected from the literature (Prochaska et al.\ 2007a; Savaglio et al.\ 2009), except for GRB\,080210 ($z=2.641$ and $\log Z/Z_\odot =-0.93$; De Cia et al.\ 2011), GRB\,081008 ($z=1.968$ and $\log Z/Z_\odot =-0.52$; D'Elia et al.\ 2011), GRB\,090926 ($z=2.106$ and $\log Z/Z_\odot =-1.89$; Rau et al.\ 2011), and GRB\,100219A ($z=4.667$ and $\log Z/Z_\odot =-1.10$; Th\"one et al.\ 2011). The dashed line is the linear correlation for QSO-DLA points. Solid curves are metallicities expected for star-forming galaxies with different stellar masses, derived by using the empirical mass-metallicity relation and its redshift evolution (Savaglio et al.\ 2005), after proper normalization at $z=0$. Metallicities are derived using Zn\textsc{ii}, Si\textsc{ii}, S\textsc{ii}, and Fe\textsc{ii} for GRB-DLAs, and Zn\textsc{ii}, Si\textsc{ii} and Fe\textsc{ii} for QSO-DLAs, after dust-depletion correction. The upper bound for G0 is set by assuming that its metallicity cannot be higher than 10 times solar. Some GRB-DLAs do not have error-bars due to lack of reported uncertainties on the measured column densities.}
\label{fZ}
\end{figure}

In Figure~\ref{fZ} we show the redshift distribution of metallicities as derived in GRB-DLAs\footnote{For more on GRB-DLA metallicities, see Berger et al.\ (2006), Fynbo et al.\ (2006), Savaglio (2006), Prochaska et al.\ (2007b) and Rau et al. (2010).} and QSO-DLAs. DLAs detected in QSO spectra are distributed between the QSO and the observer, and not directly related to the QSO itself. On the other hand, GRB-DLAs are generally associated with the galaxy hosting the burst. The average metallicity and dispersion in 17 GRB-DLAs in $z=2.0-4.5$ (excluding G0 and G1) is $\log (Z/Z_\odot) = -0.98\pm0.64$, whereas this is  $\log (Z/Z_\odot) = -0.83\pm0.76$ if we include G0 and G1. The average metallicity in 186 QSO-DLAs in the same redshift interval is $\log (Z/Z_\odot) = -1.39\pm0.61$  (Figure~\ref{fZ}). Zinc and sulfur abundances in G1 and G0 are [Zn/H] $=+0.29\pm0.10$ and [S/H] $=+0.67\pm0.34$, respectively (blue filled circles in  Figure~\ref{fZ}). Uncertainties are dominated by errors on H\,\textsc{i} columns\footnote{The upper bound to $N_{\rm H\,I}$ of G0 is artificially set to give a metallicity 10 times the solar value.}, $N_{\rm H\,I} = 10^{20.72\pm0.09}$ \cm\ and $N_{\rm H\,I} = 10^{19.62\pm0.33}$ \cm\ for G1 and G0, respectively. However, these (and thus the metallicities) are not independent of each other. A higher $N_{\rm H\,I}$ in one system would imply a lower $N_{\rm H\,I}$ in the second. If we consider one absorber for the Ly$\alpha$, we derive $N_{\rm  H\,I} = 10^{20.76\pm0.08}$ \cm\ and $z=3.5690$ (the redshift is not fixed), which is equivalent to the total HI columns of G0 plus G1 ($N_{\rm  H\,I} = 10^{20.75\pm0.09}$ \cm). From this, we get a total metallicity [Zn/H] $= +0.25\pm0.09$ and [S/H] $= +0.07\pm0.08$.
Regardless, both absorbing systems exhibit super-solar metallicities, well above QSO-DLAs and all other GRB-DLAs (Figure~\ref{fZ}). The high abundances in G0 and G1 argue against an explanation invoking a natural metal enhancement in star-forming environments associated with the birth sites of GRB progenitors (Prochaska et al.\ 2007b; Fynbo et al.\ 2008): it would be very unlikely for the two metal-rich regions, widely separated in velocity space, to be located both nearby \grb. 
%%%%%%%%%%%%
Adopting a very conservative approach, we estimate a lower limit to the metallicity by assuming that the total HI column density is in either of the two systems. These would give [Zn/H]$ > +0.25\pm0.09$ and [S/H]$ > -0.47\pm0.09$.
%%%%%%%%%%%

The highest metallicities in the local Universe are generally found in massive quiescent elliptical galaxies, for which further metallicity enrichment is prevented by the lack of cold gas, necessary for star formation. High metallicities are reached very rapidly (Matteucci 1994), which means, given the old age of the stellar population, at high redshift. The mass-metallicity relation in galaxies suggest that the typically low stellar mass of $z < 2$ GRB hosts (of the order of $\sim 10^{9.3}$ M$_\odot$) is not in contradiction with the low metallicities measured (Savaglio et al.\ 2009; Levesque et al.\ 2010). What happens at higher redshift is not totally understood. The high metallicity in G0 and G1 suggests high stellar masses. Model predictions in Figure~\ref{fZ} indicate $M_\ast > 10^{11}$ M$_\odot$ (Savaglio et al.\ 2005). Although these are based on measurements from emission lines, which indicate the average metallicity of the entire galaxy, these must be correlated with those measured in a DLA along one sight line, with some dispersion.  The stellar component where G0 and G1 have origin was detected with the multi-band imager GROND (Greiner et al.\ 2008) and the measured AB magnitudes are $r' = 24.87 \pm 0.15$, $i' = 24.25 \pm 0.18$, $J>22.7$, $H>22.2$ and $K>21.6$ (McBreen et al.\ 2010). Unfortunately, the detection in the optical only is not constraining the stellar mass because it probes the rest-frame UV, where the mass-to-light ratio can vary by a factor $> 100$. The emission (not resolved) has a size $<6$ kpc, but this is not constraining either, because galaxies (both early and late types) in the distant Universe are several times smaller than local galaxies with similar masses (Buitrago et al.\ 2008; van Dokkum et al.\ 2008; Damjanov et al.\ 2009; Toft et al.\ 2009; Mosleh et al.\ 2011). From the optical detection, we can estimate the total (integrated over G0 and G1) star-formation rate (SFR) by assuming that the rest-frame UV emission originates in the hot gas where stars form, and using the relation between the rest-frame UV flux (for instance at $\lambda = 1500$ \AA) and SFR in GRB host galaxies (Savaglio et al.\ 2009). From the observed fluxes at $z\sim3.57$, we estimate $L_{1500} = 4\times10^{40}$ erg s$^{-1}$ \AA$^{-1}$, and a non-dust corrected SFR $=6.4$ M$_\odot$ yr$^{-1}$. This value is relatively high, if one considers that the (unconstrained) dust attenuation in the galaxy can easily boost the SFR.

The G0 and G1 absorbers could be in a single massive galaxy, with high metallicity and velocity dispersion. They could also be two galaxies tidally disrupted early on, in a metal-rich environment, for instance in a galaxy cluster, whose metallicity in the intracluster medium can rise above the solar value (Sanders \& Fabian 2006). If the gas mass in G0 and G1 is not dominating over the total baryonic galaxy mass, further residual star formation would increase the metallicity over time by a small quantity. The system would eventually evolve into an elliptical galaxy after a major merger event, without dramatically affecting our view of the chemical state of the local Universe, where over-solar metallicities are possible in quiescent  $M_\ast>10^{10}$ M$_\odot$ galaxies. Alternatively, observational biases may have prevented us from detecting objects like G0 or G1 in larger numbers. Star formation proceeded very rapidly in the past, which would have led to a very rapid increase in the metallicity and in the exhaustion of gas.  What would remain today would be very metal-rich galaxies escaping observations because star formation would have ceased early on, and ten billion years later the system would contain mainly old and faint stars. 

Double absorbers are not rare in GRB afterglow spectra. We statistically studied the fraction of these strong double absorption systems with small physical separation in GRB afterglow spectra (Savaglio et al.\ 2002; Klose et al.\ 2004; Ferrero et al.\ 2009; Page et al.\ 2009) and found 5 out of 40 spectra ($\sim12$\%). For comparison, we studied a sample of about 500 QSO-DLAs, and found 18 (about 4\% of the total) close pairs (or triplets) at a distance in velocity space $< 10,000$ \kms. The excess in GRB-DLAs suggests that an enhanced star formation due to interaction and merger of galaxies could be a key mechanism triggering high-$z$ GRB events (Lopez-Sanchez \& Esteban 2009; Chen 2011), and not, as commonly assumed, the low metallicity of the progenitor (MacFadyen \& Woosley 1999). 
At $z<1$, Mannucci et al.\ (2011) suggested that the low metallicity measured in GRB hosts can be explained by the fundamental mass-metallicity relation: galaxies with similar stellar mass have higher star formation rate if more metal-poor. Campisi et al.\ (2011) studied this problem using cosmological simulations of GRB hosts and also concluded that a high SFR is likely the fundamental parameter correlated with GRBs. The task now is to seek similar indications at high redshift.
High metallicities and interactions are suggested by another class of high$-z$  galaxies\footnote{Solar and super-solar metallicities at $z>3.5$ were measured before, in the proximity of QSOs, by using narrow absorption and broad emission line systems (Srianand \& Petitjean 2000; Dietrich et al.\ 2003; Jiang et al.\ 2007). However, these systems contain 1/10 or less of the total baryonic content of their host galaxies and have a marginal effect on the global metal content.}: the rare dusty  and exceptionally luminous sub-mm galaxies (SMGs; Blain et al.\ 2002). 
Metallicities are estimated by converting the sub-millimeter emission 
(which traces molecules and dust) into mass of metals (Blain et al.\ 2002), for instance assuming, as found in the local Universe, that the dust mass is about 40\% of the total mass of metals (Franceschini et al.\ 2000). 
Metallicities derived from the nebular emission lines in SCUBA galaxies at $1.4<z<2.7$ are slightly below solar (Swinbank et al.\ 2004). A connection between some high-$z$ GRB host galaxies and SMGs is supported by the high luminosities ($L > 10^{12}$ L$_\odot$) found in 20\% of the GRB hosts observed in the radio with SCUBA (Berger et al.\ 2003). From the optical-to-radio SED of $0.8 < z < 1.5$ GRB hosts, Micha{\l}owski et al. (2008) suggest that these galaxies can be similar to the hotter (thus less luminous) sub-sample of SMGs. In GRB hosts, the dust-hidden (not detected in UV/optical) star formation  is revealed in the sub-mm, from which a more likely SFR $>100$  M$_\odot$ yr$^{-1}$ is measured (Castro Cer{\'o}n et al.\ 2006; Micha{\l}owski et al.\ 2008; Hunt et al.\ 2011). This can be a factor 10 to 100, or more, higher than measured in the UV/optical (Savaglio et al. 2009). If this applies also to G0 and G1, for which the UV emission indicates SFR $> 6$  M$_\odot$ yr$^{-1}$, a SFR higher than 100 M$_\odot$ yr$^{-1}$ is possible. The role of dust is still far from been sufficiently understood, but a better insight is now provided by early and deep SED observations of complete samples of GRB afterglows (Greiner et al.\ 2010; Kr{\"u}hler et al.\ 2011). It is found that the majority of those defined as dark GRBs were instead dimmed by  dust  along the sight lines. 

The situation might be different in the nearby Universe. In a recent work, Yates et al.\ (2011) report that mergers of massive galaxies at low redshift are much more frequent when the metallicity is low, in contrast with what is proposed here for the $z \approx 3.6$ Universe. The chemical evolution of interacting galaxies has been investigated by Perez et al.\ (2011) and Torrey et al.\ (2011) with cosmological simulations. A firm conclusion is that the star formation history in massive mergers is complex  and different from isolated galaxies. According to Torrey et al.\ (2011), gas rich systems have higher metallicity after the interaction than metal poor ones.

\section{Summary}

We report the analysis of the heavy-element enrichment of two DLAs at \zb\ and \zr\ detected in the VLT/FORS2 spectrum of the afterglow of \grb. One or both absorbers, separated in velocity space by about 660 \kms, are in the vicinity of the GRB event itself. Which one is closer is not clear. If the velocity difference is purely due to the cosmological expansion, the comoving distance between the two absorbers would be 5.8 Mpc. The blue and the red absorption systems (galaxies G1 and G0, respectively) show very strong metal lines, from which we derive zinc and sulfur over-solar abundances in the ISM:  [Zn/H] $= +0.29\pm0.10$ and [S/H] $= +0.67\pm0.34$, respectively. The two systems could be physically separated, but because of the high metal abundance in both, we propose the possibility that the two systems are interacting and in the process of merging. Ground optical detections have not enough resolution to resolve the components. Nevertheless,  the UV rest-frame image gives a high star formation rate SFR $>6$ M$_\odot$ yr$^{-1}$, not corrected for dust extinction. The double absorbers, the high metallicity at high $z$ and high SFR suggest that the two galaxies are related to massive sub-mm galaxies. If this is the case, the interaction might be one important condition through which high-$z$ GRB events occur. The situation is different from what is seen at $z<2$, where GRB hosts generally show relatively low metallicities and stellar masses. Considering that the global star formation rate of the Universe moves from big to small systems, going from high to low redshift, our discovery supports the idea that massive star formation is the main mechanism triggering GRBs and not metallicity.

The discovery of metal rich galaxies at $z>3$ is important for our understanding of the cosmic structure formation and heavy-element enrichment from a metal free era to the present time abundances. GRBs are certainly special events that might occur in special environments. However, through follow-up spectroscopy of afterglows, we can highlight regions of the Universe that are hard to detect in other ways. Most of the gas probed by our observations is perhaps in a galaxy nearby the GRB. If galaxies as those discovered by us are common, an unrecognized population of metal-rich galaxies might exist, which could  significantly alter our understanding of the cosmic chemical evolution. The challenge now is to find such remnants.

\section*{Acknowledgments}

We appreciate comments from Lise Christensen, Karl Glazebrook, Eliana Palazzi, Xavier Prochaska, and Patricia Schady, and the excellent support by the La Silla and Paranal Observatory staff, in particular C.\ Lidman. TK acknowledges support by the Deutsche Forschungsgemeinschaft Cluster of Excellence ``Origin and Structure of the Universe" as well as support by the European Commission under the Marie Curie Intra-European Fellowship Programme. The Dark Cosmology Centre is funded by the Danish National Research Foundation. SMB acknowledges support by the European Union Marie Curie European Reintegration Grant within the 7th Program under contract number PERG04-GA-2008-239176. ACU and VS acknowledge support of the Max Planck Institute for Extraterrestrial Physics.

\label{lastpage}


\begin{thebibliography}{99}
%\bibitem[Asplund et al.\ 2009]{asplund09}
%Asplund, M., Grevesse, N., Sauval, A.~J., \& Scott, P.\ 2009, ARA\&A, 47, 481
\bibitem[Berger et al. (2003)]{berger03} 
Berger, E., Cowie, L.~L., Kulkarni, S.~R., Frail, D.~A., Aussel, H., \& Barger, A.~J.\ 2003, ApJ, 588, 99 
\bibitem[Berger et al.(2006)]{berger06}
Berger, E., Penprase, B.~E., Cenko, S.~B., Kulkarni, S.~R., Fox, D.~B., Steidel, C.~C., \& Reddy, N.~A.\ 2006, ApJ, 642, 979 
\bibitem[Blain et al.\ (2002)]{blain02}
Blain, A.~W., Smail, I., Ivison, R.~J., Kneib, J.-P., \& Frayer, D.~T.\ 2002, Physics Reports, 369, 111
\bibitem[Bohlin et al.\ (1978)]{bohlin78}
Bohlin, R. C., Savage, B. D., \& Drake, J.~F.\ 1978, ApJ, 224, 132
\bibitem[Buitrago et al.(2008)]{2008ApJ...687L..61B} 
Buitrago, F., Trujillo, I., Conselice, C.~J., Bouwens, R.~J., Dickinson, M., \& Yan, H.\ 2008, ApJ, 687, L61
%\bibitem[Calzetti et al.(2000)]{2000ApJ...533..682C} 
%Calzetti, D., Armus, L., Bohlin, R.~C., Kinney, A.~L., Koornneef, J., \& Storchi-Bergmann, T.\ 2000, ApJ, 533, 682 
\bibitem[Campisi et al.(2011)]{2011arXiv1105.1378C} 
Campisi, M.~A., Tapparello, C., Salvaterra, R., Mannucci, F., \& Colpi, M.\ 2011, MNRAS, , 417, 1013
\bibitem[Castro Cer{\'o}n et al.(2006)]{2006ApJ...653L..85C} 
Castro Cer{\'o}n, J.~M., Micha{\l}owski, M.~J., Hjorth, J., Watson, D., Fynbo, J.~P.~U., \& Gorosabel, J.\ 2006, ApJL, 653, L85 
\bibitem[Cenko et al.(2010)]{cenko10} 
Cenko, S.~B., et al.\ 2011, ApJ, 732, 29
\bibitem[Cenko \& Perley (2009)]{cenko09}
Cenko, S.~B., \& Perley, D.~A.\ 2009, GRB Coordinates Network, 9027
\bibitem[Chen(2011)]{2011arXiv1110.0487C} 
Chen, H.-W.\ 2011, MNRAS, in press, arXiv:1110.0487
\bibitem[Chen et al.\ (2007)]{chen07}
Chen, H.-W., Prochaska, J.~X., Ramirez-Ruiz, E., Bloom, J.~S., Dessauges-Zavadsky, M., \& Foley, R.~J.\ 2007, ApJ,  663, 420
\bibitem[Chornock et al.\ (2009)]{chornock09}
Chornock, R., Perley, D.~A., Cenko, S.~B., \& Bloom, J.~S.\  2009, GRB Coordinates Network, 9028
%\bibitem[Cowie et al.\ 2006]{cowie96} 
%Cowie, L.~L., Songaila, A., Hu, E.~M., \& Cohen, J.~G.\ 1996,  AJ, 112, 839
\bibitem[Christensen et al.(2011)]{2011ApJ...727...73C} 
Christensen, L., Fynbo, J.~P.~U., Prochaska, J.~X., Th{\"o}ne, C.~C., de Ugarte Postigo, A., \& Jakobsson, P.\ 2011, ApJ, 727, 73 
\bibitem[Cooke et al.(2011)]{2011MNRAS.412.1047C} 
Cooke, R., Pettini, M., Steidel, C.~C., Rudie, G.~C., \& Jorgenson, R.~A.\ 2011, MNRAS, 412, 1047
\bibitem[Damjanov et al.\ (2009)]{damjanov09}
Damjanov, I., et al.\ 2009, ApJ , 695, 101
%\bibitem[Dav{\'e} et al.\ (2009)]{dave09} 
%Dav{\'e}, R., Finlator, K., Oppenheimer, B.~D., Fardal, M., Katz, N., Kere{\v s}, D., \& Weinberg, D.~H.\ 2010,  MNRAS, 404, 1355
\bibitem[De Cia et al.(2011)]{2011MNRAS.412.2229D} 
De Cia, A., et al.\ 2011, MNRAS, 412, 2229 
\bibitem[D'Elia et al.(2009)]{2009ApJ...694..332D} 
D'Elia, V., et al.\ 2009, ApJ, 694, 332 
\bibitem[D'Elia et al.(2011)]{2011arXiv1108.1084D} 
D'Elia, V., Campana, S., Covino, S., D'Avanzo, P., Piranomonte, S., \& Tagliaferri, G.\ 2011, MNRAS, in press, arXiv:1108.1084
\bibitem[Dessauges-Zavadsky et al.(2003)]{2003MNRAS.345..447D} 
Dessauges-Zavadsky, M., P{\'e}roux, C., Kim, T.-S., D'Odorico, S., \& McMahon, R.~G.\ 2003, MNRAS, 345, 447 
\bibitem[Dietrich et al.\ (2003)]{dietrich03}
Dietrich, M., Hamann, F., Shields, J.~C., Constantin, A., Heidt, J., J{\"a}ger, K., Vestergaard, M., \& Wagner, S.~J.\ 2003, ApJ, 589, 722 
\bibitem[Ferrero et al.\ (2009)]{ferrero09}
Ferrero, P., et al.\ 2009, A\&A, 497, 729
\bibitem[{Franceschini} et al. (2000)]{franceschini00}
Franceschini, A., Bassani, L., Cappi, M., Granato, G.~L., Malaguti, G., Palazzi, E., \& Persic, M.\  2000, A\&A, 353, 910
\bibitem[Fynbo et al.\ (2006)]{fynbo06}
Fynbo, J.~P.~U., et al.\ 2006, A\&A, 451, L47 
\bibitem[Fynbo et al.\ (2008)]{fynbo08}
Fynbo, J.~P.~U., Prochaska, J.~X., Sommer-Larsen, J., Dessauges-Zavadsky, M., \& M{\o}ller, P.\  2008, ApJ, 683, 321
\bibitem[Greiner et al.\ (2008)]{greiner08}
Greiner, J., et al.\ 2008, PASP, 120, 405
\bibitem[Greiner et al.(2011)]{2011A&A...526A..30G} 
Greiner, J., et al.\ 2011, A\&A, 526, A30 
%\bibitem[Howk \& Sembach 1999]{howk99}
%Howk, J.~C., \& Sembach, K.~R.\  1999, ApJ, 523, L141
\bibitem[Hunt et al.(2011)]{2011ApJ...736L..36H} 
Hunt, L., Palazzi, E., Rossi, A., Savaglio, S., Cresci, G., Klose, S., Micha{\l}owski, M., \& Pian, E.\ 2011, ApJ, 736, L36 
\bibitem[Jenkins (2009)]{jenkins09}
Jenkins, E.~B.\ 2009,  ApJ, 700, 1299
\bibitem[Jiang et al.\ (2007)]{jiang07} 
Jiang, L., Fan, X., Vestergaard, M., Kurk, J.~D., Walter, F., Kelly, B.~C., \& Strauss, M.~A.\  2007, ApJ, 134, 1150
\bibitem[Kanekar et al. (2009)]{2009ApJ...705L..40K} 
Kanekar, N., Smette, A., Briggs, F.~H., \& Chengalur, J.~N.\ 2009, ApJ, 705, L40 
\bibitem[Kennea et al.\ (2009)]{kennea09}
Kennea, J., Evans, P., \& Goad, M.\  2009, GRB Coordinates Network, 9024
\bibitem[Klose et al.\ (2004)]{klose04}
Klose, S., et al.\  2004, ApJ, 128, 1942
\bibitem[Kr{\"u}hler et al.(2011)]{2011arXiv1108.0674K} 
Kr{\"u}hler, T., et  al.\ 2011, A\&A, 534, A108
\bibitem[Ledoux et al.\ (2009)]{ledoux09}
Ledoux, C., Vreeswijk, P.~M., Smette, A., Fox, A.~J., Petitjean, P., Ellison, S.~L., Fynbo, J.~P.~U., \& Savaglio, S.\ 2009, A\&A, 506, 661 
\bibitem[Levesque et al.(2010)]{2010AJ....139..694L} 
Levesque, E.~M., Berger, E., Kewley, L.~J., \& Bagley, M.~M.\ 2010, AJ, 139, 694 
\bibitem[Lopez-Sanchez \& Esteban (2009)]{lopez-sanchez09}
Lopez-Sanchez, A.~R., \& Esteban, C.\ 2009, A\&A, 508, 615
\bibitem[MacFadyen \& Woosley (1999)]{macfadyen99}
MacFadyen, A.~I., \& Woosley, S.~E.\ 1999, ApJ , 524, 262
\bibitem[Maiolino et al.\ (2008)]{maiolino08}
Maiolino, R.\ {\it et al.} 2008, A\&A, 488, 463
\bibitem[Mannucci et al.(2009)]{2009MNRAS.398.1915M} 
Mannucci, F., et al.\ 2009, MNRAS, 398, 1915
\bibitem[Mannucci et al.(2011)]{2011MNRAS.414.1263M} 
Mannucci, F., Salvaterra, R., \& Campisi, M.~A.\ 2011, MNRAS, 414, 1263 
\bibitem[Matteucci(1994)]{1994A&A...288...57M}
Matteucci, F.\ 1994, A\&A, 288, 57 
\bibitem[McBreen et al.\ (2010)]{mcbreen10}
McBreen, S., Kr\"uhler, T., Rau, A., Greiner, J., Kann, A., Savaglio, S., {\it et al.} 2010,  A\&A,  516, A71
\bibitem[Meiring et al.\ (2008)]{meiring08}
Meiring, J.~D., Kulkarni, V.~P., Lauroesch, J.~T., P{\'e}roux, C., Khare, P., York, D.~G., \& Crotts, A.~P.~S.\ 2008, MNRAS,  384, 1015 
%\bibitem[Meiring et al.\ 2009]{meiring09}
%Meiring, J.~D., Lauroesch, J.~T., Kulkarni, V.~P., P{\'e}roux, C., Khare, P., \& York, D.~G.\ 2009, MNRAS, 397, 2037
\bibitem[Micha{\l}owski et al.(2008)]{2008ApJ...672..817M} 
Micha{\l}owski, M.~J., Hjorth, J., Castro Cer{\'o}n, J.~M., \& Watson, D.\ 2008, ApJ, 672, 817
\bibitem[Michelson (2008)]{michelson08}
Michelson, P.\ 2008, 37th COSPAR Scientific Assembly , 37, 2028
\bibitem[Molaro et al.\ (2000)]{molaro00}
Molaro, P., Bonifacio, P., Centuri{\'o}n, M., D'Odorico, S., Vladilo, G., Santin, P., \& Di Marcantonio, P.\  2000, ApJ, 541, 54
\bibitem[Mosleh et al.(2011)]{2011ApJ...727....5M} 
Mosleh, M., Williams, R.~J., Franx, M., \& Kriek, M.\ 2011, ApJ, 727, 5 
\bibitem[Nissen et al.(2007)]{2007A&A...469..319N} 
Nissen, P.~E., Akerman, C., Asplund, M., Fabbian, D., Kerber, F., Kaufl, H.~U., \& Pettini, M.\ 2007, A\&A, 469, 319
\bibitem[Nissen et al.(2004)]{2004A&A...415..993N} 
Nissen, P.~E., Chen, Y.~Q., Asplund, M., \& Pettini, M.\ 2004, A\&A, 415, 993
\bibitem[Ohno et al.\ (2009)]{ohno09}
Ohno, M., Cutini, S., McEnery, J., Chiang, J., \& Koerding, E.\ 2009, GRB Coordinates Network, 9021
\bibitem[Page et al. (2009)]{page09}
Page, K.~L., et al.\ 2009, MNRAS, 400, 134
\bibitem[Perez et al.(2011)]{2011arXiv1106.4556P} 
Perez, J., Michel-Dansac, L., \& Tissera, P.\ 2011, MNRAS, 417, 580
\bibitem[P{\'e}roux et al.(2003)]{2003MNRAS.345..480P} 
P{\'e}roux, C., Dessauges-Zavadsky, M., D'Odorico, S., Kim, T.-S., \& McMahon, R.~G.\ 2003, MNRAS, 345, 480 
\bibitem[P{\'e}roux et al.\ (2007)]{peroux07}
P{\'e}roux, C., Dessauges-Zavadsky, M., D'Odorico, S., Kim, T.-S., \& McMahon, R.~G.\ 2007, MNRAS,  382, 177
\bibitem[Pettini et al.\ (2002)]{pettini02}
Pettini, M., Rix, S.~A., Steidel, C.~C., Adelberger, K.~L., Hunt, M.~P., \& Shapley, A.~E.\ 2002, ApJ, 569, 742
\bibitem[Pettini et al.\ (2007)]{pettini97}
{Pettini}, M.\ , Smith, L.~J., King, D.~L., \& Hunstead, R.~W.\ 1997, ApJ , 486, 665
\bibitem[Prochaska(2006)]{2006ApJ...650..272P} 
Prochaska, J.~X.\ 2006, ApJ, 650, 272
\bibitem[Prochaska et al.\ (2007b)]{prochaska07}
{Prochaska}, J.\ X.\ {\it et al.} 2007b, ApJ, 666, 267
\bibitem[Prochaska et al.\ (2007a)]{prochaska07b}
{Prochaska}, J.\ X.\ {\it et al.} 2007a, ApJS, 168, 231
\bibitem[Prochaska \& Wolfe(1997)]{1997ApJ...487...73P} 
Prochaska, J.~X., \& Wolfe, A.~M.\ 1997, ApJ, 487, 73
\bibitem[Rau et al. (2010)]{rau10} 
Rau, A., et al.\ 2010, ApJ, 720, 862
%\bibitem[Sanders \& Fabian 2006]{sanders06} 
%Sanders, J.~S., \& Fabian, A.~C.\ 2006, MNRAS, 371, 1483
%\bibitem[{Rupke} {\it et al.} (2008)]{rupke08}
%Rupke, D.~S.~N., Veilleux, S., \& Baker, A.~J.\  2008, ApJ, 674, 172
\bibitem[Sanders \& Fabian (2006)]{sanders06} 
Sanders, J.~S., \& Fabian, A.~C.\ 2006, MNRAS, 371, 1483
\bibitem[Savage \& Sembach (1991)]{savage91}
{Savage}, B.\ D., \& {Sembach}, K.\ R.\ 1991, ApJ, 379, 245
\bibitem[Savaglio(2006)]{2006NJPh....8..195S} 
Savaglio, S.\ 2006, New Journal of Physics, 8, 195 
\bibitem[Savaglio et al.\ (2002)]{savaglio02}
Savaglio, S., et al.\ 2002, GRB Coordinates Network, 1633
\bibitem[Savaglio et al.\ (2005)]{savaglio05}
Savaglio, S., et al.\ 2005, ApJ, 635, 260
\bibitem[Savaglio \& Fall(2004)]{2004ApJ...614..293S} 
Savaglio, S., \& Fall, S.~M.\ 2004, ApJ, 614, 293 
\bibitem[Savaglio et al.\ (2003)]{2003ApJ...585..638S} 
Savaglio, S., Fall, S.~M., \& Fiore, F.\ 2003, ApJ, 585, 638
\bibitem[Savaglio et al.\ (2009)]{savaglio09}
Savaglio, S., Glazebrook, K., \& LeBorgne, D.\ 2009, ApJ, 691, 182
\bibitem[Schady et al.(2009)]{2009AIPC.1111..520S} 
Schady, P., et al\ 2011, A\&A, in press, arXiv:1110.3218
\bibitem[Shapley et al.\ (2004)]{shapley04}
Shapley, A., {\it et al.} 2004, ApJ, 612, 108
\bibitem[Spergel et al.(2003)]{2003ApJS..148..175S} 
Spergel, D.~N., et al.\ 2003, ApJS, 148, 175 
\bibitem[Spitzer (1978)]{spitzer78}
{Spitzer}, L.\ 1978, NewYork: Wiley
\bibitem[Srianand \& Petitjean (2000)]{srianand00}
Srianand, R., \& Petitjean, P.\ 2000, A\&A, 357, 414
\bibitem[Stratta et al.(2004)]{2004ApJ...608..846S} 
Stratta, G., Fiore, F., Antonelli, L.~A., Piro, L., \& De Pasquale, M.\ 2004, ApJ, 608, 846 
\bibitem[{Swinbank} et al. (2004)]{swinbank04}
Swinbank, A.\ M. {\it et al.}, 2004, ApJ, 617, 64
\bibitem[Th{\"o}ne et al.(2011)]{2011AN....332..281T} 
Th{\"o}ne, C.~C., Fynbo, J., Goldoni, P., de Ugarte Postigo, A., Covino, S., Campana, S., \& X-shooter GRB collaboration 2011, Astronomische Nachrichten, 332, 281 
\bibitem[Toft et al.\ (2009)]{toft09}
Toft, S., Franx, M., van Dokkum, P.~G., Forster-Schreiber, N.~M., Labbe, I., Wuyts, S., \& Marchesini, D. 2009, ApJ, 705, 255
\bibitem[Torrey et al.(2011)]{2011arXiv1107.0001T} 
Torrey, P., Cox, T.~J., Kewley, L., \& Hernquist, L.\ 2011, arXiv:1107.0001
\bibitem[{Updike} {\it et~al.} (2009)]{updike09}
Updike, A.~C., Filgas, R., Kr\"uhler, T., Greiner, J., \& McBreen, S.\ 2009, GRB Coordinates Network, 9026
\bibitem[van Dokkum et al.\ (2008)]{vandokkum08}
van Dokkum, P.~G., et al.\ 2008, ApJ, 677, L5
\bibitem[Vladilo et al.(2011)]{2011A&A...530A..33V} 
Vladilo, G., Abate, C., Yin, J., Cescutti, G., \& Matteucci, F.\ 2011, A\&A, 530, A33
\bibitem[Vreeswijk et al.(2004)]{2004A&A...419..927V} 
Vreeswijk, P.~M., et al.\ 2004, A\&A, 419, 927 
\bibitem[Vreeswijk et al.\ (2007)]{vreeswijk07}
Vreeswijk, P.~M., et al.\ 2007,  A\&A, 468, 83
\bibitem[Wolfe et al.\ (2003)]{wolfe03}
Wolfe, A.~M., Gawiser, E., \& Prochaska, J.~X.\ 2003, ApJ, 593, 235
\bibitem[Wolfe et al.(2005)]{2005ARA&A..43..861W} 
Wolfe, A.~M., Gawiser, E., \& Prochaska, J.~X.\ 2005, ARA\&A, 43, 861
\bibitem[Yates et al.(2011)]{2011arXiv1107.3145Y} 
Yates, R.~M., Kauffmann, G., \& Guo, Q.\ 2011, MNRAS, submitted, arXiv:1107.3145
\bibitem[Zafar et al.(2011)]{2011A&A...532A.143Z} 
Zafar, T., Watson, D., Fynbo, J.~P.~U., et al.\ 2011, A\&A, 532, A143
\end{thebibliography}
\end{document}